\documentclass[a4paper,11pt]{article}
\usepackage{cite}
\usepackage{fullpage,setspace,hyperref,comment,caption}
\usepackage[font=footnotesize,labelfont=bf,margin=1cm]{caption}
\usepackage{cite} 
\usepackage{color}
\usepackage{amsfonts}
\usepackage{amsmath}
\usepackage{amssymb}
\usepackage{todonotes} 
\usepackage[utf8]{inputenc}
\usepackage{slashed}
\usepackage{graphicx}
\usepackage{pstricks-add}
\usetikzlibrary{arrows,snakes,backgrounds}
\newcommand{\be}{\begin{equation}}
\newcommand{\ee}{\end{equation}}

\begin{document}
\begin{titlepage}
\vfill
\begin{flushright}
\end{flushright}
\vfill

\begin{center}
   \baselineskip=16pt
   	{\Large \bf Field theory aspects of $ \eta $-deformed superstring background}
   	\vskip 2cm
 	{\large \bf  Dibakar Roychowdhury$^a$\footnote{\tt dibakar.roychowdhury@ph.iitr.ac.in} }
   	\vskip .6cm
   	{\it  $^a$ Department of Physics, Indian Institute of Technology Roorkee, \\ Roorkee 247667, Uttarakhand, India \\ \ \\}
   	\vskip 2cm
\end{center}

\begin{abstract}
We explore various field theory aspects of integrable $ \eta $-deformed geometry in type IIB supergravity by employing several holographic probes. These include the computation of holographic timelike entanglement entropy and estimation of various other field theory observables for example, the flow central charge and the quantum complexity. We also discuss the associated brane set up and compute Page charges. We further use them to calculate the coupling constant in the dual QFTs considering both small and large deformation limits.
\end{abstract}

\vfill

\setcounter{footnote}{0}
\end{titlepage}
\tableofcontents 
\section{Overview and motivation}
The integrable $ \eta $-deformation \cite{Delduc:2013qra}-\cite{Delduc:2014kha} of various AdS supergravity backgrounds had witnessed an overwhelming and ever growing interest for the last decade due to its several novel aspects. The $ \eta $- deformed backgrounds\footnote{This is a generalisation of the $ \eta $-deformed PCM \cite{Klimcik:2002zj}-\cite{Klimcik:2008eq} and the symmetric space model of \cite{Delduc:2013fga}.} are primarily of interest because the associated sigma model is integrable \cite{Delduc:2013qra}-\cite{Delduc:2014kha}, \cite{Roychowdhury:2017vdo}. These backgrounds are solutions of type IIB supergravity equations \cite{Arutyunov:2013ega}-\cite{Jimbo:1985zk} and satisfy the unimodularity condition \cite{Borsato:2016ose}.

Based on these ideas, the authors of \cite{Hoare:2018ngg} considered a new class of R-matrices that satisfy the unimodularity condition when all the simple roots of the Dynkin diagram are fermionic. These R-matrices correspond to a particular $ q $-deformation of the (super)isometry algebra \cite{Delduc:2013fga}, \cite{Delduc:2016ihq}. While the stringy counterpart of the duality is well established, the dual QFTs living on the holographic screen \cite{Kameyama:2014vma}-\cite{Kameyama:2014via} is much less explored. 

\paragraph{Goal and general theme of this paper.} Let us first elaborate on the general theme of this paper and give a broader perspective of different pieces of calculations that are carried out. As we shall see, all these results are well connected under a central theme, which we elaborate now. 

The big picture here is to understand the field theory aspects \cite{Roychowdhury:2017oed}-\cite{Roychowdhury:2017oqd} of $ \eta $-deformed $AdS_2 \times S^2 \times T^6$ backgrounds in a holographic set up. Here, we are primarily interested in the following questions - (i) How do we compute various field theory observables in a QFT that is dual to $ \eta $-deformed $AdS_2 \times S^2 \times T^6$ and (ii) Is there a way to realise these QFTs as a world-volume theory living inside a D-brane, much similar in spirit as that of a typical Hanany-Witten like brane construction \cite{Hanany:1996ie}. The first part of this paper is devoted towards a full understanding of various QFT observables, which include the holographic entanglement entropy \cite{Ryu:2006bv}-\cite{Hubeny:2007xt}, the central charge \cite{Macpherson:2014eza}-\cite{Chatzis:2024kdu} and the complexity \cite{Susskind:2014rva}-\cite{Fatemiabhari:2024aua}. Since the dual QFT is ($0+1$)d, therefore only boundary coordinate relevant to our analysis is the (bulk) global time in  Lorentzain signature. In other words, the entanglement entropy in the boundary theory corresponds to the (holographic) timelike entanglement  \cite{Doi:2022iyj}-\cite{Roychowdhury:2025ukl} in the context of ($0+1$)d QFTs. As our analysis reveals, timelike entanglement entropy (tEE) can be related to other field theory observables \cite{Roychowdhury:2025ukl}, namely the flow central charge \cite{Macpherson:2014eza}-\cite{Chatzis:2024kdu} and the complexity \cite{Susskind:2014rva}-\cite{Fatemiabhari:2024aua}, in a holographic RG flow, where the tip of the extremal surface can be tuned from deep IR upto the UV cut-off in the bulk.

In the second part of this paper, we argue that the dual ($0+1$)d QFT could be engineered using D-branes and in particular can be recast as a defect within a ($1+3$)d QFT living inside the D3 brane world-volume. We show this explicitly by constructing the associated Page charge, which reveals a non-zero value for the D3 brane. We further show that the world-volume theory effectively boils down into a ($0+1$)d QFT, which becomes strongly coupled in the near boundary limit. Combining the above pieces together, we argue that the QFT dual to $\eta$-deformed $AdS_2 \times S^2 \times T^6$ background could be portraied as a defect living inside D3 brane world-volume for which one should be able to compute various field theory observables.

\paragraph{The background.} Before proceeding further, it is customary to make a brief review on the $\eta$-deformed background that is relevant for our analysis. We begin by reviewing the $ \eta $-deformed $ AdS_2 \times S^2 \times T^6 $ background in type IIB \cite{Hoare:2018ngg}, which is given by the following line element
\begin{align}
\label{e1}
&ds_{10}^2 = ds^2_4 + ds^2_{T^6}\\
&ds^2_4= \frac{L^2}{(1-\kappa^2 \rho^2)}\Big(-(1+ \rho^2)dt^2+\frac{d\rho^2}{1+ \rho^2} \Big)+\frac{L^2}{(1+ \kappa^2 r^2)}\Big((1-r^2)d \phi^2 +\frac{dr^2}{1-r^2}  \Big)\\
&ds^2_{T^6} =L^2 d\varphi_i d\varphi_i ~;~ \kappa = \frac{2 \eta}{1- \eta^2}~;~i=4, \cdots , 9
\label{e3}
\end{align}
where $ \kappa $ is the (real) deformation parameter that ranges between $0 $ to $ \infty $.

The above background \eqref{e1}-\eqref{e3} is supported by a dilaton of the form
\begin{align}
\label{e4}
e^{-2 \Phi}=e^{-2 \Phi_0}\frac{(1 - \kappa^2 \rho^2)(1+ \kappa^2 r^2)}{1-\kappa^2(\rho^2 - r^2 - \rho^2 r^2)}.
\end{align}

In what follows, we adopt the following change of coordinates
\begin{align}
\rho= \sinh\varrho~;~ r = \cos \theta
\end{align}
and rewrite the metric  \eqref{e1}-\eqref{e3} and the dilaton  \eqref{e4} in global coordinates as
\begin{align}
\label{e6}
&ds^2_4= \frac{L^2}{(1-\kappa^2 \sinh^2 \varrho)}\Big(-\cosh^2\varrho dt^2+d \varrho^2 \Big)+\frac{L^2}{(1+ \kappa^2 \cos^2 \theta)}\Big(d\theta^2+\sin^2 \theta d\phi^2  \Big)\\
&e^{-2 \Phi}=e^{-2 \Phi_0}\frac{(1 - \kappa^2 \sinh^2\varrho)(1+ \kappa^2 \cos^2\theta)}{1-\kappa^2( \sinh^2\varrho \sin^2\theta - \cos^2\theta)}.
\label{e7}
\end{align}

The first term in \eqref{e6} is the deformed $ AdS_2 $ where $ \varrho $ is the global radial coordinate. As in the undeformed case, $ \varrho=\varrho_{IR} =0 $ corresponds to the deep IR limit in the bulk. On the other hand, the UV limit corresponds to setting $ \Lambda_{UV}=\sinh^{-1}\Big(\frac{1}{\kappa} \Big) $, which clearly goes to infinity in the limit of the vanishing deformation. We identify $  \Lambda_{UV} $ as the UV cut-off in the bulk, that sets the location of the holographic screen \cite{Kameyama:2014vma}-\cite{Kameyama:2014via} where the dual ($ 0+1 $)d QFT is living.

\paragraph{Organisation.} The organisation for the rest of the paper is as follows. In section 2, we revisit the RT prescription \cite{Ryu:2006bv}-\cite{Hubeny:2007xt} of computing entanglement entropy and in particular, the recent proposal of computing timelike entanglement entropy (tEE) \cite{Doi:2022iyj}-\cite{Afrasiar:2024ldn}. We provide an RG interpretation of our findings in terms of holographic $ c $ function \cite{Macpherson:2014eza}-\cite{Chatzis:2024kdu} and complexity \cite{Susskind:2014rva}-\cite{Fatemiabhari:2024aua}, that we compute in the following section 3. Next, in section 4, we explore different Page charges and the associated brane set up \cite{Lozano:2020txg}. We also calculate the coupling constant in the dual QFTs by probing the geometry with D branes and in particular following the lines of \cite{Lozano:2020txg}-\cite{Lozano:2016kum}. Finally, we summarise all our findings in section 5 and conclude the paper along with some future remarks and key observations.
\section{Timelike entanglement}
In the original Ryu-Takayanagi (RT) approach \cite{Ryu:2006bv}-\cite{Hubeny:2007xt}, the bulk radial coordinate emerges naturally through the quantum entanglement in the boundary QFT living on a spacelike hypersurface. On a similar spirit, as argued by authors in \cite{Doi:2023zaf}, the global time in the bulk $AdS$ emerges through quantum entanglement in the boundary QFT living on timelike hypersurface. The extremal surface for holographic timelike entanglement (in Lorentzian signature) is a union of spacelike and timelike geodesics, that contains a turning point in the bulk. This is in contrary to the previously defined RT prescription \cite{Ryu:2006bv}-\cite{Hubeny:2007xt}, where the extremal surface is spacelike. 

We would like to extend the above ideas by performing an explicit calculation on tEE \cite{Doi:2023zaf} for $\eta$- deformed  $AdS_2 \times S^2 \times T^6$ backgrounds (where we work in the Lorentzian signature) and to give it an interpretation in terms of other field theory observables. Since the boundary theory is ($ 0+1 $)d, therefore quantum entanglement in these QFTs corresponds to integrating out timelike subregions. On the dual gravitational counterpart, this corresponds to considering a class of extremal surfaces in the bulk that ends on a timelike surface in the boundary \cite{Doi:2022iyj}-\cite{Afrasiar:2024ldn}, which for the present case is the location of the holographic screen. 

To begin with, we propose $ t=t(\varrho) $, which results in a co-dimension one surface
\begin{align}
ds^2_9 = (g_{\varrho \varrho }-g_{tt}t'^2)d \varrho^2 +g_{\theta \theta}d\theta^2 + g_{\phi \phi}d\phi^2+g_{\varphi_i \varphi_i}d\varphi_i^2
\end{align}
where we identify the individual metric coefficients as
\begin{align}
\label{e9}
&g_{tt}=\frac{L^2 \cosh^2\varrho}{(1-\kappa^2 \sinh^2 \varrho)}~;~g_{\varrho \varrho}=\frac{L^2}{(1-\kappa^2 \sinh^2 \varrho)}\\
&g_{\theta \theta}=\frac{L^2}{(1+ \kappa^2 \cos^2\theta)}~;~g_{\phi \phi}=\frac{L^2 \sin^2\theta}{(1+ \kappa^2 \cos^2\theta)}~;~g_{\varphi_i \varphi_i}=L^2.
\label{e10}
\end{align}

The timelike entanglement entropy (tEE) is defined as the area of the co-dimension one hyper-surface, weighted by the dilaton
\begin{align}
\label{e11}
\mathcal{S}^{(tEE)}=\frac{V_{T}}{4G_N} \int_{S^2_\kappa} d\theta  d\phi \sqrt{\det g_{2}} \int_{0}^{\Lambda_{UV}}d\varrho e^{-2 \Phi} \sqrt{g_{\varrho \varrho} -g_{tt}t'^2}
\end{align}
where $ V_T $ represents the volume of the six torus ($ T^6 $) and $ \det g_2 = g_{\theta \theta}g_{\phi \phi}$ is the determinant of the metric on the deformed unit two sphere ($ S^2_\kappa (\theta , \phi)$).

The equation of motion for $ t(\varrho) $ yields
\begin{align}
\label{e12}
t'^2(\varrho)=\frac{C^2 g_{\varrho \varrho}}{g_{tt}(C^2+e^{-4 \Phi} g_{tt})}
\end{align}
where $ C $ is the constant of integration.

Substituting \eqref{e12} into \eqref{e11}, one finds
\begin{align}
\label{e13}
\mathcal{S}^{(tEE)}=\frac{V_{T} }{4G_N}\int_{S^2_\kappa} d\theta d \phi \sqrt{\det g_{2}} \int_{0}^{\Lambda_{UV}}d\varrho  \frac{e^{-4 \Phi}\sqrt{g_{tt}g_{\varrho \varrho}}}{\sqrt{C^2 +e^{-4 \Phi} g_{tt}}}.
\end{align}
\subsection{Small deformation limit}
Expanding the integrand in \eqref{e13}, in the limit of small deformation ($ \kappa \ll 1 $), we find
\begin{align}
\label{e14}
\mathcal{S}^{(tEE)}&=\frac{\pi V_{T} L^4}{G_N} \int_{0}^{\infty}d\varrho \frac{e^{-4 \Phi_0}\cosh \varrho }{\sqrt{C^2+L^2 e^{-4 \Phi_0 } \cosh ^2\varrho }}\nonumber\\
&-\frac{\pi \kappa^2  V_{T} L^4}{4G_N} \int_0^\pi d \theta \sin \theta \int_{0}^{\infty}d\varrho \frac{e^{-4 \Phi_0}\cosh \varrho \Gamma (\theta , \varrho)}{\sqrt{C^2+L^2 e^{-4 \Phi_0 } \cosh ^2\varrho }}
\end{align}
where we denote the function $\Gamma (\theta , \varrho)$, appearing in \eqref{e14}, as
\begin{align}
\Gamma (\theta , \varrho)& = \frac{1}{(C^2 e^{4 \Phi_0 }+L^2 \cosh ^2\varrho)}\Big[  2 C^2 e^{4 \Phi_0 } \left(\cos 2 \theta \sinh ^2\varrho +\cos ^2\theta \right)\nonumber\\&+L^2 \cosh ^2\varrho \left(\cos 2 \theta \cosh ^2\varrho +1\right)\Big].
\end{align}

Performing the $ \theta $ integral, one finds
\begin{align}
&\mathcal{S}^{(tEE)}=\frac{\pi V_{T} L^4}{G_N} \int_{0}^{\infty}d\varrho \frac{e^{-4 \Phi_0}\cosh \varrho }{\sqrt{C^2+L^2 e^{-4 \Phi_0 } \cosh ^2\varrho }}\Big( 1+ \frac{\kappa^2 }{12}\bar{\Gamma} ( \varrho)\Big)\\
&\bar{\Gamma} ( \varrho)=\frac{2 C^2 e^{4 \Phi_0 } (\cosh 2 \varrho -3)+L^2 \cosh ^2\varrho  (\cosh 2 \varrho -5)}{ \left(C^2 e^{4 \Phi_0 }+L^2 \cosh ^2\varrho \right)}.
\end{align}

In order to find the turning point in the bulk one sets $ C=i \tilde{C} $, such that $ C^2 = -\tilde{C}^2<0 $ \cite{Afrasiar:2024lsi}. On top of that, we introduce the following change of variables
\begin{align}
\label{e18}
\cosh \varrho = \frac{1}{z}
\end{align}
such that, $ z=1 $ stands for the deep IR limit in the bulk. On the other hand, $ z_{UV}=\epsilon \ll 1$ corresponds to the UV cut-off in the bulk. Using this change of variable \eqref{e18}, one obtains
\begin{align}
\label{e19}
\mathcal{S}^{(tEE)}=\frac{\pi V_{T} L^4}{G_N} \int_{\epsilon}^{1}\frac{dz}{z^2}\frac{e^{-4 \Phi_0 }}{\sqrt{1-z^2}}\frac{1}{\sqrt{\frac{L^2 }{z^2}e^{-4 \Phi_0 }-\tilde{C}^2}}\Big( 1+ \frac{\kappa^2 }{12}\bar{\Gamma} ( z)\Big).
\end{align}

Clearly, the turning point ($ z=z_0 $) is obtained by setting $ \tilde{C}=\frac{L}{z_0}e^{-2\Phi_0}$, where $ z_0 <1 $. The next step is to substitute it into \eqref{e19}, which finally reveals $ \mathcal{S}^{(tEE)}=-i z_0 \mathcal{S}^{(\text{Im})}$ where
\begin{align}
\label{e20}
&\mathcal{S}^{(\text{Im})}=\frac{2\pi V_{T} L^3}{G_N} \int_{z_0}^{1}\frac{dz}{z}\frac{e^{-2 \Phi_0 }}{\sqrt{1-z^2}\sqrt{z^2-z^2_0}}\Big( 1+ \frac{\kappa^2 }{12}\bar{\Gamma} ( z)\Big)\\
&\bar{\Gamma} ( z)=\frac{-8 z^4+6 z^2 z_0^2+4 z^2-2 z_0^2}{3 z^2(z^2- z_0^2)}.
\end{align}

Let us check asymptotic properties of \eqref{e20}. To begin with, we push the tip ($ z_0 $) of the extremal surface near UV, $ z_0 \sim \epsilon \sim 0$. This yields the following expression for the tEE \eqref{e20}
\begin{align}
\label{e22}
|\mathcal{S}^{(\text{Im})}|_{z_0 \sim \epsilon}&=\frac{2\pi V_{T} L^3}{G_N}e^{-2 \Phi_0 }\int_\epsilon^1 dz \frac{\kappa ^2+\left(9-2 \kappa ^2\right) z^2}{9 z^4 \sqrt{1-z^2}}\nonumber\\
&=\frac{2\pi \kappa^2 V_{T} L^3}{27 G_N \epsilon^3}e^{-2 \Phi_0 }\Big(  1 + \frac{9 \epsilon^2}{\kappa^2}(6- \kappa^2)+\cdots\Big)
\end{align}
which clearly exhibits a pole of order $ 3 $, in limit $ \epsilon \rightarrow 0 $. Comparing with the undeformed ($ \kappa =0 $) example, which reveals a pole of order one, one might therefore conclude that the tEE diverges at a faster rate (near the asymptotic infinity) in the presence of non zero $ \eta $-deformation.  

Using \eqref{e22}, we finally note down 
\begin{align}
\label{e23}
\mathcal{S}^{(tEE)}|_{z_0 \sim \epsilon}=-\frac{2\pi i \kappa^2 V_{T} L^3}{27 G_N \epsilon^2}e^{-2 \Phi_0 }\Big( 1+ \mathcal{O}(\epsilon^2/\kappa^2)\Big)
\end{align}
which clearly exhibits a pole of order $2$ near asymptotic infinity ($ \epsilon \rightarrow 0 $). As we show in the subsequent section, this is precisely the pole structure appearing in the holographic flow central charge ($c_{flow}$) near UV. Combing them, it is quite suggestive to argue that tEE is measure of degrees of freedom for ($ 0+1 $)d QFTs that are dual to $ \eta $-deformed superstring backgrounds.

On the other hand, considering a deep IR ($ z_0 \sim 1 $) limit in the bulk, one finds
\begin{align}
|\mathcal{S}^{(\text{Im})}|_{z_0 \sim 1}= \frac{\pi V_{T} L^3}{18 G_N}e^{-2 \Phi_0 }\Big[ \frac{\kappa^2}{z^2_0}(1-z^2_0)-3(6 - \kappa^2)\log \Big( \frac{\Lambda_{IR}z^2_0}{1-z^2_0}\Big)\Big]_{z_0 \sim 1}
\end{align}
where, $ \Lambda_{IR}\sim 0 $ is a small number such that the ratio $ \frac{\Lambda_{IR}z^2_0}{1-z^2_0} $ is finite and $ \mathcal{O}(1) $ in the deep IR limit. The first term, on the other hand, vanishes identically as we approach $ z_0 \sim1 $.

Combining the above facts together, one might therefore argue that tEE \eqref{e20} decreases in a RG flow from UV ($ z_0 \sim 0 $) to deep IR ($ z_0 \sim 1 $). This precisely reflects the properties of a $ c $ function \cite{Macpherson:2014eza}-\cite{Chatzis:2024kdu} in a RG flow, which we further justify in the subsequent section. The UV ($ z_0 \sim \epsilon \sim 0 $) divergence is an artefact of the large area contribution due to the extremal surface. 

The corresponding subsystem length at the boundary turns out to be
\begin{align}
\label{e25}
t_{sub}=2\int_0^1 \frac{dz}{\sqrt{1-z^2}}=\pi.
\end{align}

The above formula \eqref{e25} denotes the subsystem size corresponding to the largest possible extremal surface, that could be constructed in the bulk. In pother words, \eqref{e25} represents the maximal area of the entangling region (which is a timelike strip for the present example), in the boundary ($0+1$)d QFT, living on a timelike hypersurface. 
\subsection{Comments on large deformation limit}
Let us briefly comment about the large $\kappa(\gg 1) $ regime, which corresponds to a holographic screen located at a finite radial cut-off, $\Lambda_{UV}=\frac{1}{\kappa}+\mathcal{O}(\kappa^{-3})$ in the bulk. The associated metric functions \eqref{e9}-\eqref{e10} and the background dilaton \eqref{e7} read as
\begin{align}
&g_{tt}=-\frac{L^2 \coth ^2\varrho}{\kappa ^2}~;~g_{\varrho \varrho}=-\frac{L^2 \text{csch}^2\varrho}{\kappa ^2}\\
&g_{\theta \theta}=\frac{L^2 \sec ^2\theta }{\kappa ^2}~;~g_{\phi \phi}=\frac{L^2 \tan ^2\theta}{\kappa ^2}~;~g_{\varphi_i \varphi_i}=L^2\\
&e^{-2 \Phi}=-e^{-2 \Phi_0}\Big[ \frac{\kappa ^2 }{\text{csch}^2\varrho -\tan ^2\theta }\nonumber\\
& - \frac{ \left(\sin ^2\theta  \sinh ^2\varrho \left(\sinh ^2\varrho -\cos ^2\theta \right)+\cos ^4\theta \right)}{\left(\cos ^2\theta -\sin ^2\theta  \sinh ^2\varrho \right)^2} \Big].
\end{align}

Considering a large $ \kappa $ expansion of the entities in \eqref{e13}, we find at LO in $ 1/\kappa $
\begin{align}
\mathcal{S}^{(tEE)}=\frac{-i\pi V_{T} L^3}{2G_N \kappa}e^{-2 \Phi_0}\int_0^\pi \frac{\sin \theta}{\cos^2 \theta} d \theta \int_{0}^{1/\kappa}\frac{d\varrho }{\sinh \varrho}\frac{1}{\left(\text{csch}^2\varrho -\tan ^2\theta \right)}
\end{align}
which after performing the $ \theta $ integral independently yields
\begin{align}
\label{e30}
\mathcal{S}^{(tEE)}|_{\kappa \gg 1}&=\frac{i\pi V_{T} L^3}{4G_N \kappa}e^{-2 \Phi_0}\int_{0}^{1/\kappa}d\varrho   \left(\log \left(e^{-3 \varrho }\right)-\log \left(e^{\varrho }\right)\right) \text{sech}\varrho  \nonumber\\
&=-\frac{i\pi V_{T} L^3}{2G_N \kappa^3}e^{-2 \Phi_0}+\cdots.
\end{align}

Clearly, tEE vanishes when the deformation is large enough ($ \kappa \gg 1 $). The vanishing of the tEE simply follows from the impossibility of constructing an extremal surface (that ends on a timelike subregion) in the limit of large deformation ($ \kappa \gg 1 $). One should think of this as an artifact of a limiting situation, in which the holographic screen is pushed deep inside the bulk and therefore the size of the extremal surface goes to zero.
\section{QFT observables}
The purpose of this section is to calculate various observables in the dual QFT, that count the number of degrees of freedom in a holographic RG flow from UV asymptotic to deep IR inside the bulk geometry. These are generally denoted in terms of the flow central charge \cite{Chatzis:2024kdu} and the complexity \cite{Susskind:2014rva}-\cite{Fatemiabhari:2024aua}, which we explain below. Our goal would be to find an interrelation between these entities and the tEE that has been estimated in the previous section. 
\subsection{Flow central charge}
For generic QFTs, the $ c $ function or flow central charge ($ c_{flow}(\varrho)$) counts the number of degrees of freedom that is monotonically decreasing in a RG flow, namely $\frac{d c_{flow}}{d \varrho}|>0$, where $ 0< \varrho < \infty $. In the case of a RG flow from a UV fixed point (at $\varrho=\infty$) to a IR fixed point (at $\varrho=0$), the flow function ($ c_{flow}(\varrho) $) captures the corresponding central charges such that, $ c_{UV}>c_{IR} $, where $c_{UV}=c_{flow}(\varrho)|_{\varrho=\infty}$ and $c_{IR}=c_{flow}(\varrho)|_{\varrho=0}$.

The calculation that we carry out, precisely follows the algorithm developed by authors in a series of papers \cite{Roychowdhury:2025ukl}-\cite{Chatzis:2024kdu}. To begin with, we re-express the type IIB background as
\begin{align}
&ds^2_{10}=\frac{L^2}{(1-\kappa^2 \sinh^2 \varrho)}\Big(-\cosh^2\varrho dt^2+d \varrho^2 \Big)+\mathcal{G}_{ab}d \omega^a d \omega^b\\
&\mathcal{G}_{ab}d \omega^a d \omega^b=\frac{L^2}{(1+ \kappa^2 \cos^2 \theta)}\Big(d\theta^2+\sin^2 \theta d\phi^2  \Big)+L^2 d \varphi_i \varphi_i
\end{align}
where $ \mathcal{G}_{ab} $ is the metric of the eight dimensional internal manifold.

For a ($ 0+1 $)d QFT, the flow central charge is given by the volume of the internal eight manifold, weighted by the dilaton
\begin{align}
\label{e33}
c_{flow}(\varrho)&=\frac{1}{G_N}\int d^8\omega e^{-2\Phi}\sqrt{\det \mathcal{G}_{ab}}\nonumber\\
&=\frac{2 \pi V_T}{G_N}\int_0^\pi d \theta e^{-2\Phi}\sqrt{\det g_2}.
\end{align}

Considering a small $ \kappa (\ll 1) $ expansion of the integrand in \eqref{e33}, we find
\begin{align}
c_{flow}(\varrho)=\frac{2 \pi V_T L^2}{G_N}e^{-2\Phi_0}\int_0^\pi d \theta \sin\theta  (1 - \kappa^2 \cos ^2\theta \cosh ^2\varrho).
\end{align}

Performing the $ \theta $ integral and using \eqref{e18}, we find
\begin{align}
\label{e35}
|c_{flow}(z)|=\frac{4 \pi V_T L^2}{G_N}e^{-2\Phi_0}\Big| 1-\frac{\kappa^2}{3 z^2} \Big|.
\end{align}

As we show below, the flow central charge \eqref{e35} diverges near the UV aymptotic ($ z \sim \epsilon \sim 0 $) of the bulk. On the other hand, it clearly decreases as we RG flow from the UV asymptotic into the deep IR ($ z \sim 1 $) of the bulk. Combining the above facts together, we can claim that \eqref{e35} actually counts the number of degrees of freedom of the dual QFT, in a holographic RG flow.

 Taking a UV limit of \eqref{e35}, we find
\begin{align}
\label{e36}
|c_{flow}(z)|_{z \sim \epsilon}=\frac{4 \pi \kappa^2 V_T L^2}{3G_N \epsilon^2}e^{-2\Phi_0}
\end{align}
which reveals an identical pole structure as found in \eqref{e23}.

Comparing \eqref{e23} and \eqref{e36}, we map these entities near the UV asymptotic
\begin{align}
\label{e37}
\mathcal{S}^{(tEE)}|_{UV}=-\frac{i L}{18}|c_{flow}|_{UV}.
\end{align}

The above relation \eqref{e37} is quite intriguing, which shows that near the UV asymptotic of the bulk, the tEE is approximately equal to the flow function in the dual QFT. In other words, the UV degrees of freedom (in the dual QFT) can be captured using the notion of timelike entanglement along a timelike subregion, which addresses one of the basic themes as posited in the introduction, namely the mapping between tEE and the central charge in the dual QFT. 
\subsection{Complexity}
Complexity is yet another measure of number of degrees of freedom in a QFT. These degrees of freedom are represented by number of elementary gates that are required to build up a quantum circuit. In what follows, in our analysis, we follow the CV proposal that was outlined and successfully implemented in a series of papers \cite{Chatzis:2024kdu}-\cite{Fatemiabhari:2024aua}. The basic idea is to compute the volume of a nine manifold weighted by the dilaton, that is defined at a fixed boundary time.

We express the 10d metric \eqref{e1} in the following form
\begin{align}
ds^2_{10}=-g_{tt}dt^2 + ds^2_9.
\end{align}

Following the holographic prescription \cite{Chatzis:2024kdu}-\cite{Fatemiabhari:2024aua}, complexity in the dual QFT is defined as the maximal volume associated with the spacelike hyper-surface at a constant time, $ t=t_0 $
\begin{align}
\label{e39}
\mathcal{C}_V& = \frac{1}{G_N}\int d^9x e^{-2\Phi}\sqrt{\det g_9}\nonumber\\
&=c_{flow}\int_0^{\Lambda_{UV}}d\varrho \sqrt{g_{\varrho \varrho}}\nonumber\\
&=\frac{L}{8}c_{flow} \mathcal{F}(\Lambda_{UV})
\end{align}
where we denote the above function $\mathcal{F}(\Lambda_{UV})$ as
\begin{align}
 \mathcal{F}(\Lambda_{UV})=\frac{\kappa^2 }{2}e^{2 \Lambda_{UV}}-2 \left(\kappa ^2-4\right) \Lambda_{UV}.
\end{align}

Using \eqref{e37} and \eqref{e39}, we finally relate tEE and complexity in the UV limit of dual QFT
\begin{align}
\label{e41}
\mathcal{C}_V = \frac{9i}{4}\mathcal{S}^{(tEE)}|_{UV} \mathcal{F}(\Lambda_{UV}).
\end{align}

The above relation \eqref{e41} along with \eqref{e37} relates various field theory observables with tEE near the UV asymptotic of the $\eta$-deformed background. To summarise, tEE plays an alternative definition of complexity or the central charge in the UV limit of the dual QFT. 
\subsection{Remarks about large $ \kappa $ deformation}
Let us now investigate above entities namely the complexity and the flow central charge in the limit of large deformations ($ \kappa \gg 1 $). Expanding the integrand \eqref{e33} in the large $ \kappa(\gg 1) $ regime, we obtain the flow central charge as
\begin{align}
c_{flow}=-\frac{2 \pi V_T L^2}{G_N }e^{-2\Phi_0}\int_0^\pi d \theta  \frac{\sin \theta \sinh^2 \varrho}{(\cos ^2\theta -\sin ^2\theta  \sinh ^2\varrho )}
\end{align}
which by virtue of the $ \theta $ integral yields
\begin{align}
\label{e43}
c_{flow}|_{\varrho \sim \Lambda_{UV}}=\frac{4 \pi V_T L^2}{G_N \kappa^2 }e^{-2\Phi_0}.
\end{align}

Clearly, \eqref{e43} is vanishingly small at LO in $ 1/\kappa $ expansion. In fact, there is no notion of RG flow in the large deformation limit. This stems from the fact that the holographic screen is pushed well inside the bulk interior and the corresponding RG flow sizes to exist.  

A close comparison between \eqref{e30} and \eqref{e43} reveals
\begin{align}
\label{e44}
\mathcal{S}^{(tEE)}|_{\kappa \gg 1}=-\frac{i L}{8 \kappa}c_{flow}|_{\varrho \sim \Lambda_{UV}}.
\end{align}

On a similar note, we estimate complexity \eqref{e39} in the limit of large deformation
\begin{align}
\mathcal{C}_V|_{\kappa \gg 1}&=c_{flow}\int_{\Lambda_{IR}}^{1/\kappa}d\varrho \sqrt{g_{\varrho \varrho}}\nonumber\\
&=\frac{L}{\kappa}c_{flow}\Big( \log \left(\frac{1 }{2 \kappa \Lambda_{IR}}\right)-\frac{1}{12 \kappa^3 }\Big)
\end{align}
where $ \Lambda_{IR}\sim 0 $ is some IR cut-off in the bulk such that $ \kappa \Lambda_{IR} \sim \mathcal{O}(1) $.

Finally, comparing with \eqref{e44}, we relate complexity with tEE near the UV limit
\begin{align}
\label{e46}
\mathcal{C}_V|_{\kappa \gg 1}=8i \mathcal{S}^{(tEE)}|_{\kappa \gg 1}\Big( \log \left(\frac{1 }{2 \kappa \Lambda_{IR}}\right)-\frac{1}{12 \kappa^3 }\Big).
\end{align}

The above results \eqref{e44} and \eqref{e46} indicate that the mapping between different QFT observables and the tEE persists even in the regime of strong deformations ($\kappa \gg 1$), irrespective of the fact that these entities are now vanishingly small. The smallness of these entities follows from the fact that the bulk degrees of freedom are largely reduced as a result of the displacement of the holographic screen deep inside the bulk.
\section{Page charges and brane set up}
We now move onto the second part of our project, where we engineer the dual one dimensional QFT using D-branes. As mentioned before, we interpret this one dimensional QFT as a defect within a four dimensional QFT describing the world-volume physics of a D3 brane. To understand the one dimensional QFT as a theory living on D brane, the first step is to construct the quantised Page charges associated with background RR fluxes. It is the flux of background RR forms over the internal cycles of the $\eta$-deformed background that yields different types of D branes that can be associated with the supergravity solution \eqref{e1}. These background RR fluxes are constructed using $ R $-matrix that satisfies unimodularity condition \cite{Hoare:2018ngg}.

Let us briefly summarise what we find below. The $\eta$-deformed supergravity background possesses non zero Page charges for the D5 and D3 brane and yields a zero Page charge for NS5 branes. One should think of these D3 brane world-volume directions wrapping the three cycles of the six torus ($ T^6 $), that are localised along the bulk radial direction ($\rho = \rho_0$). On a similar token, D5 brane world-volume directions wrap the five cycles of the six torus ($T^6$) and are expanded along $R_t \times T^5$, as shown in the Table below. The dual QFT is what we identify as 1d defect \cite{Lozano:2021rmk} living inside a 4d QFT describing D3 branes. However, such an interpretation breaks down in the large deformation ($ \kappa \gg 1 $) limit, in which D3 branes sizes to exist.

\begin{center}
\begin{tabular}{||c c c c c c c c c c c||} 
 \hline
Dp & t & $\rho$ &$ r$ & $\phi$ & $\varphi_4$ & $ \varphi_5 $ & $ \varphi_6 $ & $ \varphi_7 $ & $ \varphi_8 $ & $ \varphi_9 $ \\  [0.5ex] 
 \hline\hline
 D3 & x &  &  & & x &x & x &  & &    \\ 
 \hline
 D5 & x &  &  & & x &x & x & x &x &   \\ [1ex] 
 \hline
\end{tabular}
\end{center}
\subsection{Background RR fluxes and Page charges}
The RR three form field strength associated with $\eta$-deformed background \eqref{e1} is given by \cite{Hoare:2018ngg} 
\begin{align}
\label{e47}
\mathcal{F}_3 &= -\mathcal{N}\kappa r (1+ \kappa^2 r^2)d\rho \wedge(d \varphi_4 \wedge d \varphi_5+d \varphi_6 \wedge d \varphi_7+d \varphi_8 \wedge d \varphi_9)\nonumber\\
&-\mathcal{N}\kappa \rho (1- \kappa^2 \rho^2)dr \wedge(d \varphi_4 \wedge d \varphi_5+d \varphi_6 \wedge d \varphi_7+d \varphi_8 \wedge d \varphi_9)\nonumber\\
&+\mathcal{N}\kappa \rho (1-r^2)d \rho \wedge dt \wedge d\phi - \mathcal{N}\kappa r (1+ \rho^2)dr \wedge dt \wedge d\phi
\end{align}
where the pre-factor $\mathcal{N}$ can be expressed as
\begin{align}
\mathcal{N}=\frac{\sqrt{1+ \kappa^2}}{\sqrt{1- \kappa^2 \rho^2}\sqrt{1+ \kappa^2 r^2}(1- \kappa^2(\rho^2 -r^2 - \rho^2 r^2))}.
\end{align}

The Page charge counts the number of Dp branes that the background RR fluxes are coupled to. Given a ($ 8-p $) background RR flux $ \mathcal{F}_{8-p}  $, the Page charge can be expressed as \cite{Lozano:2020txg}-\cite{Lozano:2016kum}
\begin{align}
\mathcal{Q}_{Dp}=\frac{1}{(2 \pi)^{7-p}}\int_{\Sigma^{(8-p)}} \mathcal{F}_{8-p}.
\end{align}

Setting $ t= $ constant in \eqref{e47}, the number of D5 branes turns out to be
\begin{align}
\label{e50}
\mathcal{Q}_{D5}=\frac{1}{(2 \pi)^{2}}\int_{\Sigma^{(3)}} \mathcal{F}_{3}.
\end{align}

Considering a small $ \kappa(\ll 1) $ expansion of the three form flux \eqref{e47}, we find 
\begin{align}
\label{e51}
\mathcal{Q}_{D5}&=-\frac{3\kappa}{(2 \pi)^{2}}\Big[\int r d\rho \times (2\pi)^2+\int \rho dr \times (2\pi)^2 \Big]+\mathcal{O}(\kappa^3)\nonumber\\
&=-3\kappa \int d(\rho r)+\mathcal{O}(\kappa^3)\nonumber\\
&=3\kappa \rho_{max}+\mathcal{O}(\kappa^3)\nonumber\\
&=3+\mathcal{O}(\kappa^3)
\end{align}
where we set, $ \rho_{max}=\frac{1}{\kappa}\gg 1 $ as the location of the holographic screen. Since the Page charge counts the number of D-branes, therefore it has to be an integer, which is clearly reflected in the above expression \eqref{e51}. 

RR five form field strength, on the other hand, is given by \cite{Hoare:2018ngg} 
\begin{align}
\label{e52}
\mathcal{F}_5 &=\mathcal{N}((1+ \kappa^2 r^2)d\rho - \kappa^2 \rho r (1+ \rho^2)dr)\wedge dt \wedge\text{Re}\Omega_3\nonumber\\
&-\mathcal{N}(\kappa^2 \rho r(1-r^2)d\rho+(1- \kappa^2 \rho^2)dr)\wedge d\phi \wedge \text{Im}\Omega_3
\end{align}
where $ \Omega_3 = (d\varphi_4 - i d\varphi_5)\wedge (d\varphi_6 - i d\varphi_7)\wedge (d\varphi_8 - i d\varphi_9) $. This clearly reveals that\footnote{While integrating over $\text{Im}\Omega_3$, it contains three cycles each of which ranges between $ 0\leq \varphi_i \leq 2 \pi $, thereby contributing an overall factor of $ (2\pi)^3 $.}
\begin{align}
\text{Im}\Omega_3&=-d \varphi_4 \wedge (d \varphi_6 \wedge d \varphi_9 +d \varphi_7 \wedge d \varphi_8)\nonumber\\
&-d \varphi_5 \wedge (d \varphi_6 \wedge d \varphi_8 - d \varphi_7 \wedge d \varphi_9).
\end{align}

Setting $ t= $ constant in \eqref{e52}, we find the Page charge associated with D3 branes as
\begin{align}
\label{e54}
\mathcal{Q}_{D3}&=\frac{1}{(2 \pi)^4}\int_{\Sigma^{(5)}} \mathcal{F}_{5}\nonumber\\
&=-\frac{1}{(2 \pi)^4}\Big[ \kappa^2 \int \rho r (1-r^2)d \rho\nonumber\\
&+\int \Big(1+\frac{1}{2} \kappa ^2 \left(\rho ^2-\left(2 \rho ^2+3\right) r^2+1\right)\Big )dr \Big]\int_0^{2 \pi} d \phi \int \text{Im}\Omega_3\nonumber\\
&=-\frac{4}{(2 \pi)^4}\Big[-2+\frac{\kappa^2}{2} \int d \Big( r(\rho^2 +1)(1-r^2) \Big)+\frac{\kappa^2}{2}\int \rho^2 r^2 dr\Big] \times (2\pi)^4\nonumber\\
&=8-2 \kappa^2 \rho^2_0\int_{1}^{-1}r^2 dr\nonumber\\
&=8+\frac{4}{3}\kappa^2 \rho^2_0
\end{align}
where D3 branes are considered to be localised along the bulk radial direction, $ \rho=\rho_0 $. The radial locations $ \rho_0 $ are discrete numbers, such that the entity \eqref{e54} is an integer.

Finally, we note down possibilities for any NS5 branes in the background. The corresponding Page charge can be expressed as the flux of background NS-NS three form 
\begin{align}
\mathcal{Q}_{NS5}=\frac{1}{(2\pi)^2}\int_{\Sigma^{3}}\mathcal{H}_3=0
\end{align}
which vanishes identically by virtue of background NS-NS two form as found in \cite{Hoare:2018ngg} .

To summarise, we have a system of D3/D5 brane that couples with the $\eta$-deformed background \eqref{e1}. Taking the ratio between the number of D5 and D3 brane, one finds $\frac{Q_{D5}}{Q_{D3}}<1$. In other words, in the small deformation limit ($\kappa \ll 1$), the physics is mostly dominated by D3 branes. As we show below, in this small deformation limit, one can use the D3 brane world-volume  theory to describe an \emph{effective} one dimensional field theory, which we identify as the QFT dual to $\eta$-deformed geometry \eqref{e1}, for which we compute various field theory observables.
\subsection{D3 brane and coupling}
We begin by considering probe D3 brane in the background that is located at a fixed radial distance $ \rho = \rho_0 $ and wrapping the three cycle of the six torus ($ T^6 $). The corresponding DBI action could be formally expressed as
\begin{align}
\mathcal{S}_{D3}=T_{D3}\int d^4x \Big[ e^{-\Phi} \sqrt{-\det (g+ 2 \pi \alpha' F_2)}-\mathcal{C}_4\Big]
\end{align}
where $ T_{D3}=\frac{1}{(2 \pi)^3 g_s \alpha'^2} $. The background RR potential relevant to our analysis is given by \cite{Hoare:2018ngg} 
\begin{align}
\mathcal{C}_4=\frac{ L^4 e^{-\Phi_0} \rho_0 \sqrt{1+\kappa^2}}{\sqrt{1-\kappa^2 (\rho_0^2 - r^2 -\rho_0^2 r^2)}}dt \wedge \text{Re}\Omega_3.
\end{align}

Turning on a world-volume field strength tensor $F_2= F_{t \varphi}dt \wedge d \varphi $, that is considered to be small enough and thereby allowing a Taylor exapnsion of the world-volume action \cite{Lozano:2016kum}, yields
\begin{align}
\label{e57}
 e^{-\Phi} \sqrt{-\det (g+ 2 \pi \alpha' F_2)}-\mathcal{C}_4\simeq\frac{3 \pi \alpha'^2 }{2\rho_0}\mathcal{Q}_{D3}e^{-\Phi_0}F_{t \varphi}^2
\end{align}
where, we set $ r= \pm 1 $ and take the boundary limit $ \rho_0 \gg 1 $, such that the number of D3 branes $N_{D3}= \mathcal{Q}_{D3}\simeq \frac{4}{3}\kappa^2 \rho^2_0 $ is kept finite in the limit, $\kappa \ll 1$. 

Using \eqref{e57} and integrating over the three cycles of the six torus, we finally obtain
\begin{align}
\mathcal{S}_{D3}=\frac{3 \pi \mathcal{Q}_{D3}}{2g_s \rho_0}e^{-\Phi_0}\int dt (\partial_t \Phi)^2 =\frac{1}{g^2_{eff}}\int dt (\partial_t \Phi)^2
\end{align}
where $ g_s $ is the string coupling and $ g^2_{eff} $ is the effective coupling of ($0+1$)d QFT that is living inside the D3 brane world-volume. Here, we identify $ A_\varphi =\Phi $ as the scalar of the reduced (effective) one dimensional theory living inside the D3 brane world-volume. 

Clearly, in the near boundary limit $ \rho_0 \rightarrow \infty $, the effective one dimensional theory becomes strongly coupled ($ g^2_{eff} \sim \rho_0 \rightarrow \infty $), which we identify as the ($ 0+1 $)d QFT that is dual to $ \eta $- deformed $ AdS_2 \times S^2 \times T^6 $ supergravity background. In what follows, we identify this reduced QFT as one dimensional defect within a four dimensional QFT inside D3 brane world-volume. 
\subsection{Comments on large deformation limit}
As a final remark, we note down background RR fluxes in the large deformation ($\kappa \gg 1$) limit. Notice that, in the large deformation limit, the radial coordinate is bounded between $ 0< \rho < \rho_{max} $ where $ \rho_{max}\sim \frac{1}{\kappa}\sim 0 $. In other words, $ \kappa \rho < 1 $ for the interval considered. On the other hand, one must consider $ \kappa r \gg 1 $, given the range $ -1 \leq r \leq 1 $.

Considering above facts, we note down components of background RR fluxes that are relevant for the computation of Page charges. To begin with, we note down the three form flux
\begin{align}
\mathcal{F}_3 &= -\kappa d\rho \wedge(d \varphi_4 \wedge d \varphi_5+d \varphi_6 \wedge d \varphi_7+d \varphi_8 \wedge d \varphi_9)+\mathcal{O}(1/\kappa)
\end{align}
which yields the Page charge associated with D5 branes as
\begin{align}
\mathcal{Q}_{D5}\Big|_{\kappa \gg 1}=-\frac{1}{(2 \pi)^{2}}\int_{\Sigma^{(3)}} \mathcal{F}_{3}=3\kappa \int_0^{1/\kappa}d \rho =3.
\end{align}
In other words, the number of D5 branes is preserved. This stems from the fact that the RR  three form flux increases linearly with the deformation parameter ($ \sim \kappa $), while on the other other hand, the radius decreases inversely as $ \sim \frac{1}{\kappa} $, keeping the product finite.

On a similar note, for the five form flux we find (for constant $ t $)
\begin{align}
\mathcal{F}_5 = \Big(1- \frac{1}{r^2}\Big) \rho d\rho \wedge d \phi \wedge \text{Im}\Omega_3 +\mathcal{O}(1/\kappa^2)
\end{align}
which yields a vanishing flux in the limit of large deformations
\begin{align}
\mathcal{Q}_{D3}\Big|_{\kappa \gg 1}=-\frac{1}{(2 \pi)^4}\int_{\Sigma^{(5)}} \mathcal{F}_{5}=\frac{2}{\kappa^2} \Big(1- \frac{1}{r^2}\Big) \simeq 0.
\end{align}

In other words, the number of D3 branes vanishes, which is consistent with the fact that the dual QFT looses all its degrees of freedom in the large deformation limit. D5 brane, on the other hand, wraps the six torus ($T^6$) and does not feel the radial shrinking. To summarise, one might therefore conclude that $ \eta $- deformed backgrounds possess well defined dual description as long as the deformation parameter is small enough. In the limit of large deformations, the superstring theory looses its dual interpretation in the form QFTs living in ($ 0+1 $)d.
\section{Concluding remarks}
We now summarise the key findings of the paper. In the first part of the paper, we compute timelike entanglement entropy (tEE) in the dual QFT$ _1 $ both in the limit of small ($\kappa \ll 1$) as well as large deformations ($\kappa \gg 1$). Since the boundary theory is ($ 0+1 $)d, it is therefore expected that tEE is the only possible way of defining the entanglement between different (timelike) subsystems in the dual QFT$ _1 $. As our analysis reveals, for small deformations, the imaginary component of tEE precisely captures the notion of the flow central charge ($ c_{flow} $), in the sense that it decreases in a RG flow from UV to deep IR. In other words, the imaginary tEE is a measure of the number of degrees of freedom in the dual QFT$ _1 $ living on the holographic screen. This result should follow from dimension reduction of a parent 2d QFT, where the flow central charge in ($ 0+1 $)d would correspond to either of the left or right $ c $ functions of the parent QFT$ _2 $.

To be precise, we map these entities near the UV asymptotic \eqref{e37}. This follows quite naturally from the unique pole (of order $2$) structure appearing in both the entities near the UV asymptotic. On a similar token, one could also relate holographic complexity and tEE near the UV scale of the dual QFT$ _1 $ \eqref{e41}. Combining all these three pieces together, we conclude that tEE ($ \mathcal{S}^{(tEE)} $), flow central charge ($ c_{flow} $) and complexity ($ \mathcal{C}_V $) are all equally good measures of the number of degrees of freedom for the boundary QFT$ _1 $ near its UV limit.

We also explore the physics of dual QFT$ _1 $ in the limit of large deformations ($ \kappa \gg 1 $). It turns out that all three entities namely tEE ($ \mathcal{S}^{(tEE)} $), flow central charge ($ c_{flow} $) and complexity ($ \mathcal{C}_V $) become vanishingly small. This stems from the fact that in the large deformation limit, the holographic screen is pushed well inside the bulk interior. As a result, one does not have much access to the bulk degrees of freedom as they are enormously reduced. This is reflected in the corresponding QFT observables that actually count these degrees of freedom. In this limit, the dual QFT$ _1 $ exhausts almost all its degrees of freedom and becomes non-dynamical.

We extend our analysis one step further by computing Page charges using background NS-NS and RR fluxes. Our analysis reveals the existence of D3 and D5 branes and no NS5 branes as a part of full type IIB solution. This sort of completes the RR sector of $ \eta $- deformed type IIB $ AdS_2 $ superstrings. The question that remains to be be answered is how to engineer a ($ 0+1 $)d quantum mechanical model using these brane set up. We portray these QFT$ _1 $ as 1d defect inside a 4d QFT living within D3 brane world-volume. As a preliminary support to our conjecture, we compute the DBI action for $N_{D3}$ probe D3 branes near the boundary, which eventually boils down into a ($0+1$)d QFT living inside D3 brane world-volume. We further show that the effective coupling of this reduced theory  becomes very large indicating a strongly coupled theory.  These could be further confirmed by computing various other QFT observables/probes for example, the Wilson line and t'Hooft loops \cite{Lozano:2021rmk}, which we leave for future investigation.

\paragraph {Acknowledgements :}
The author is indebted to the authorities of IIT Roorkee for their unconditional support towards researches in basic sciences. The author also acknowledges The Royal Society, UK for financial assistance. The author also acknowledges the Mathematical Research Impact Centric Support (MATRICS) grant (MTR/2023/000005) received from ANRF, India.

\end{document}